\documentclass[a4paper]{jpconf}
\usepackage[cp1252]{inputenc}
\usepackage{iopams}
\usepackage{graphicx}
\begin{document}
	\title{Solution of the  'sign problem' in pair approximation}
	
	\author{A~S~Larkin$^1,^2$, V~S~Filinov$^1$ and V~E~Fortov$^1$}
	
	\address{$^1$Joint Institute for High Temperatures of the Russian Acad\-emy of Sciences, Izhorskaya 13 Bldg~2, Moscow 125412, Russia}
	\address{$^2$Moscow Institute for Physics and  Technology, Institutskiy per. 9, Dolgoprudny, Moscow Region, 141701,Russia}
	
	\ead{vladimi\_filinov@mail.ru}
	
	\begin{abstract}
		The main difficulty for path integral Monte Carlo studies of Fermi systems results from the requirement of antisymmetrization of the density matrix
		and is known in literature as the 'sign problem'.
		To overcome this issue the new numerical version of the %{\large {\tiny {\Huge }}}
		Wigner approach to quantum mechanics for treatment thermodynamic properties of degenerate systems of fermions has been developed. The new path integral
		representation of quantum Wigner function in the phase space has been obtained for canonical ensemble. Explicit analytical expression of the Wigner function accounting for Fermi statistical effects by effective pair pseudopotential has been proposed. Derived pseudopotential depends on coordinates, momenta and degeneracy parameter of fermions and takes into account Pauli blocking of fermions in phase space. The new quantum Monte-Carlo method for calculations
		of average values of arbitrary quantum operators has been proposed.
		To test the developed approach calculations of the momentum distribution functions and pair correlation functions 
		of the degenerate ideal system of Fermi particles has been carried out in a good agreement with analytical 
		distributions. Generalization of this approach for studies influence of interparticle interaction on 
		 momentum distribution functions of strongly coupled 
		Fermi system is in progress.
	\end{abstract}

	\section{Introduction}
	Over the last decades significant progress has been observed in theoretical studies of
	thermodynamic properties of strongly correlated fermions at non-zero temperatures,
	which is mainly conditioned by the application of numerical simulations (see review
	\cite{Ceprl1}). The reason for this success is the possibility of an explicit representation of the density matrix in the form
	of the Wiener path integrals \cite{Feynm} and making use of the Monte Carlo method for further calculations. The main difficulty for path integral Monte Carlo (PIMC) studies
	of Fermi systems results from the requirement of antisymmetrization of the density matrix
	\cite{Feynm}. As result all thermodynamic quantities are presented as the sum of alternating sign terms related to even and odd permutations and are equal to the small difference of two large numbers, which are the sums of positive and negative terms. The numerical calculation in this case is severely hampered. This difficulty is known in the literature as the 'sign problem'.
	
	To overcome this issue some approaches have been developed \cite{Ceprl1,CPP,NJP,PRL1,PRL2,Ceprl2,Ceprl3,Militz}. For example 
	approaches \cite{CPP,PRL1,PRL2} get ideal Fermi gas with very good accuracy. For interacting fermions the first results  \cite{PRL1} 
	have been obtained for the high densities for 33 particles and were now extrapolated to the thermodynamic limit very accurately in \cite{PRL2}. 
	
	The 'fixed-node method' \cite{Ceprl1,Ceprl2,Ceprl3,Militz} is very widely known. 
	However the main result of work \cite{ancntrd} is that
	the 'fixed--node method' can not reproduce even the well known
	ideal fermion density matrix and should
	be considered as an uncontrolled empirical approach for treatment thermodynamics of fermions.
	The analogous contradictions have been analytically obtained many years ago in \cite{cluster} from virial decomposition of the many fermion 'fixed-node' density matrix'.
	
	In this work to treat the 'sign problem' the new numerical version of the Wigner approach to quantum mechanics allowing studies of 
	thermodynamic properties of the degenerate systems of fermions has been developed. The new path integral
	representation of quantum Wigner function in
	the phase space has been obtained for canonical ensemble. Explicit analytical expression of the Wigner function accounting for Fermi statistical effects by effective pair pseudopotential has been proposed. Derived pseudopotential depends
	on coordinates, momenta and degeneracy parameter of fermions and takes into account Pauli blocking of fermions in phase space.
	The new quantum Monte-Carlo method for calculations of average values of arbitrary quantum operators has been proposed. To test the developed approach calculations of the momentum distribution functions of the ideal system of Fermi particles has been carried out. Calculated by Monte Carlo method the momentum distributions and 
	pair correlation functions for degenerate ideal fermions are in a good agreement with analytical distributions. Generalization of this approach for studies influence of interparticle interaction on momentum distribution functions of strongly coupled 
	Fermi system is in progress. First results show appearance of long quantum 'tails' in the Fermi distribution functions. 
	
	\section{Wigner function for canonical ensemble}
	Average value of arbitrary quantum operator $\hat A$ can be written
	as its Weyl's symbol $A(p,x)$, averaged over phase space with the
	Wigner function $W(p,x;\beta)$ \cite{Wgnr,Tatr}:
	\begin{eqnarray}\label{wigfunc_average0}
		\langle \hat A\rangle = \int\frac{dp \rmd x} {2\pi\hbar}
		A(p,x) W(p,x;\beta),
	\end{eqnarray}
	where the Weyl's symbol of operator $\hat A$ is:
	\begin{eqnarray}\label{wigfunc_weylsymbol}
		A(p,x) = \int\frac{d\xi}{2\pi\hbar}e^{-i\langle \xi|p\rangle /\hbar}
		\langle x-\xi/2|\hat A|x + \xi/2\rangle.
	\end{eqnarray}
	Weyl's symbols for usual operators like
	$\hat p$, $\hat x$, $\hat p^2$, $\hat x^2$, $\hat H$, $\hat H^2$
	etc. can be easily calculated directly from definition
	(\ref{wigfunc_weylsymbol}).
	The Wigner function of many particle system in canonical ensemble is
	defined as a Fourier transform of the off--diagonal matrix element of the
	density matrix operator in coordinate representation:
	\begin{eqnarray}\label{pathint_wignerfunction}
		W(p,x;\beta) = Z(\beta)^{-1}\int{d\xi}
		e^{i\langle p|\xi\rangle /\hbar}
		\langle x-\xi/2|e^{-\beta\hat H}|x + \xi/2\rangle.
	\end{eqnarray}
	Here $\rho=\exp(-\beta \hat{H})$ is the density matrix operator
	of a quantum
	system of particles with the Hamiltonian $\hat{H}=\hat{K}+\hat{U}$
	equal to the sum of kinetic $\hat{K}$ and potential energy $\hat{U}$
	operators, while $\beta = 1/k_B T$, $Z(\beta)$ is partition function. 
	
	There are well known difficulties in derivation of exact explicit analytical expression for Wigner
	function as operators of kinetic and potential energy in Hamiltonian do
	not commutate. To overcome this obstacle let
	us represent Wigner function in the form of path integral like
	in the well known case of the partition function. As example let us consider
	equilibrium a 3D two-component mass asymmetric electron--hole mixture consisting of
	$N_e$ electrons and $N_h$ holes ($N_h=N_e=N$) \cite{ElHol}. Here $Z(\beta)$ is defined as:
	%\cite{Wigner,NormanZamalin,feynman-hibbs}:
	\begin{equation}\label{q-def}
		Z(N_e,N_h,V;\beta) = \frac{1}{N_e!N_h! \lambda_e^{3N_e} \lambda_h^{3N_h} } \sum_{\sigma}\int\limits_V
		dx \,\rho(x, \sigma;\beta), 
	\end{equation}
	where $\rho(x, \sigma;\beta)$ denotes the diagonal matrix
	elements of the density operator ${\hat \rho} = e^{-\beta {\hat H}}$. 
	%, and $\beta=1/k_BT$.
	In equation~(\ref{q-def}), $x=\{x_e,x_h\}$ and $\sigma=\{\sigma_e,\sigma_h\}$
	are the spatial coordinates and spin degrees of freedom
	of the electrons and holes, i.e.
	$x_a=\{x_{1,a}\ldots x_{l,a}\ldots x_{N_a,a}\}$
	and $\sigma_a=\{\sigma_{1,a}\ldots \sigma_{t,a}\ldots
	\sigma_{N_a,a}\}$, $\lambda_a=\sqrt{\frac{2\pi\hbar\beta}{m_a}}$ is the thermal wave length, $l,t=1,\dots ,N_a$ with $a=e,h$. 
	
	Of course, the density matrix elements of interacting quantum
	systems is not known (particularly for low temperatures and high
	densities), but it can be constructed using a path integral
	approach~\cite{Feynm,Zamalin} based on the operator identity
	$e^{-\beta {\hat H}}= e^{-\epsilon {\hat H}}\cdot
	e^{-\epsilon {\hat H}}\dots e^{-\epsilon {\hat H}}$,
	where $\epsilon = \beta/M$, which allows us to
	rewrite the integral in equation~(\ref{q-def}) as
	\begin{eqnarray}
		&&\sum_{\sigma} \int\limits \rmd x^{(0)}\,
		\rho(x^{(0)},\sigma;\beta) =
		%\nonumber\\ &&
		\int\limits \rmd x^{(0)} \dots \rmd x^{(m)} \dots
		dx^{(M-1)} \, \rho^{(1)}\cdot\rho^{(2)} \, \dots \rho^{(M-1)} \times
		%\nonumber\\
		\nonumber\\
		&&\sum_{\sigma}\sum_{P_e} \sum_{P_h}(\pm 1)^{\kappa_{P_e}+ \kappa_{P_h}} \,
		{\cal S}(\sigma, {\hat P_e}{\hat P_h} \sigma^\prime)\, %\times
		%\nonumber\\ &&
		{\hat P_e} {\hat P_h}\rho^{(M)}\big|_{x^{(M)}= x^{(0)}, \sigma'=\sigma} \,.
		\label{rho-pimc}
	\end{eqnarray}
	In equation~(\ref{rho-pimc}) the index $m=0,\dots, M-1$
	labels the off--diagonal high-temperature density matrices
	$\rho^{(m)}\equiv \rho\left(x^{(m)},x^{(m+1)};\epsilon \right) =
	\langle x^{(m)}|e^{-\epsilon {\hat H}}|x^{(m+1)}\rangle$.
	
	With the error of order $1/M^2$
	arising from neglecting commutator $\epsilon^2/2 \left[K,U\right]$ in
	$e^{-\epsilon {\hat H}} \approx e^{-\epsilon {\hat K}} e^{-\epsilon {\hat U}} e^{-\epsilon^2/2 \left[K,U\right]} \dots$
	each high temperature factor can be presented in the form
	$\langle x^{(m)}|e^{-\epsilon {\hat H}}|x^{(m+1)}\rangle \approx
	\langle x^{(m)}|e^{-\epsilon {\hat K}}|x^{(m+1)}\rangle \langle x^{(m)}|e^{-\epsilon {\hat U}}|x^{(m)}\rangle$.
	In the limit $M\rightarrow \infty$ the error of the whole
	product of high temperature factors is equal
	to zero $(\propto 1/M)$ and this approach
	gives exact path integral representation of the partition function.
	
	The spin gives rise to the spin part of the density matrix (${\cal
		S}$) with exchange effects accounted for by the permutation
	operators $\hat P_e$ and $\hat P_h$ acting on the electron and
	hole coordinates $x^{(M)}$ and spin projections $\sigma'$. The
	sum is over all permutations with parity $\kappa_{P_e}$ and
	$\kappa_{P_h}$. So thermodynamic values are equal to very small
	difference between large (of order $N!/2$) positive and
	negative contributions giving by the 
	even and odd permutations. The problem of accurate calculation
	of this difference is the {\it{ well known sign problem for
			degenerate Fermi systems}}. The aim of this work is to develop
	simple and accurate approach for calculation this difference.
	%to a great extent to overcome the
	
	\section{Exchange effects in pair approximation}
	
	To explain the basic ideas of our approach is enough to consider
	the simple system of ideal fermions (electrons and holes),
	so further $\hat{U}\equiv0$.
	The hamiltonian of the system (${\hat H}={\hat K}={\hat K_e}+{\hat K_h}$)
	contains kinetic energy of electrons ${\hat K_e}$ and holes ${\hat K_h}$
	respectively. Due to the commutativity of these operators the
	path integral representation of density matrix (\ref{rho-pimc}) is exact at any
	finite number $M$. For our purpose it is enough to consider the sum over permutations in
	pair approximation at $M=1$ (see \cite{Huang}): 
	\begin{eqnarray}
		&&\sum_{\sigma}\sum_{P_e} \sum_{P_h}(\pm 1)^{\kappa_{P_e}+ \kappa_{P_h}} \,
		{\cal S}(\sigma, {\hat P_e}{\hat P_h} \sigma_{a}^\prime)\, %\times
		{\hat P_e} {\hat P_h}\rho \,
		\big|_{x^{1}= x^{(0)}, \sigma'=\sigma} %=\,
		\nonumber\\ &&
		=\sum_{\sigma_e}\sum_{P_e}(\pm 1)^{\kappa_{P_e}}{\cal S}(\sigma_e, {\hat P_e} \sigma_{e}^\prime)\,
		\rho_e \big|_{x_e^{1}= x_e^{(0)}, \sigma_e'=\sigma_e} \,
		\nonumber\\ &&
		\times \sum_{\sigma_h}\sum_{P_h}(\pm 1)^{\kappa_{P_h}}{\cal S}(\sigma_h, {\hat P_h} \sigma_{h}^\prime)\,
		\rho_h \big|_{x_h^{1}= x_h^{(0)}, \sigma_h'=\sigma_h} %= \,
		\nonumber\\ &&
		=\sum_{\sigma_e}\left \{ 1-\sum_{l<t}f^2_{e;lt}+\sum_{l,t,c}f_{e;lt}f_{e;lc}f_{e;tc}-\dots \right \}
		\nonumber\\ &&
		\times \sum_{\sigma_h}\left \{ 1-\sum_{l<t}f^2_{h;lt}+\sum_{l,t,c}f_{h;lt}f_{h;lc}f_{h;tc}-\dots \right \} \approx\,
		\nonumber\\ &&
		\approx\,\sum_{\sigma_e} \prod_{l<t}(1-f^2_{e;lt})\sum_{\sigma_h}\prod_{l<t}(1-f^2_{h;lt}) =
		\sum_{\sigma}\exp (-\beta \sum_{l<t}\tilde{v}^{e}_{ lt})\exp (-\beta \sum_{l<t}\tilde{v}^{h}_{ lt})
		\label{rho-fermi}
	\end{eqnarray}
	where  
	\begin{eqnarray}
		&&f_{a;lt}=\exp (-\frac{\pi|x^{(0)}_{l,a}-x^{(0)}_{t,a}|^2}{\lambda_a^2}) 
		\nonumber\\ && 
		\tilde{v}^{a}_{ lt}=-kT \ln(1-f_{a;lt}f_{a;tl})=-kT \ln(1-\delta_{\sigma_{l,a}\sigma_{t,a}}
		%\delta_{\sigma_l^a\sigma_{l}^b}
		\exp (-\frac{2\pi|x^{(0)}_{l,a}-x^{(0)}_{t,a}|^2}{\lambda_a^2}))
		\label{pairfermi}
	\end{eqnarray}
	is exchange potential \cite{Huang}. This formula shows that the
	first corrections accounting for the antisymmetrization of the
	density matrix result in the endowing particles by the pair exchange
	potential $\tilde{v}^a_{ lt}$.  Below to take into account exchange
	effects in Wigner functions we are going to use analogous pair
	potential depending on the phase space variables. 
	%---
	\section{Path integral representation of Wigner function}
	
	Antisymmetrized Wigner function can be written in the form:
	%where $P(p,x)$ is the Weyl's symbols of the antisymmetrization operator:
	\begin{eqnarray}\label{permut}
		&&W(p,x;\beta) = \frac{1}{Z(\beta)N_e!N_h!\lambda_e^{3N_e} \lambda_h^{3N_h}}
		\sum_{\sigma}\sum_{P_e} \sum_{P_h}(\pm 1)^{\kappa_{P_e}+
			\kappa_{P_h}} {\cal S}(\sigma, {\hat P_e}{\hat P_h} \sigma^\prime)
		\big|_{\sigma'=\sigma}\,
		\nonumber \\&&
		%%\times \int\frac{d \xi }{(2\pi)^{6N}}e^{i\langle \xi |p\rangle }
		\times \int d \xi e^{i\langle \xi |p\rangle }
		\langle x-\xi/2| \prod_{m=0}^{M-1}
		e^{-\epsilon {\hat K_{m}}}
		%\nonumber \\&&
		|{\hat P_e} {\hat P_h}(x + \xi/2)\rangle \,
		% \nonumber\\&&
		%=\int \rmd x^{(M_1+1)} \langle x^{(M_1)}|e^{-\tilde{\epsilon} {\hat %K}}|x^{(M_1+1)}\rangle
		%\sum_{s=1}^{N_e}\sum_{k=1}^{N_h} C^s_{N_e} C^k_{N_h} \,
		%{\rm det} \,||\psi||_{\rm sk},
		\label{rho_s}
	\end{eqnarray}
	%where $2N$ is number of particle in 3D electron -- hole plasma.   
	
	Now replacing intermediate variables of integration $x^{(m)}$ in (\ref{rho_s}) (see (\ref{rho-pimc}))
	for any permutation $P_eP_h$:
	\begin{eqnarray}\label{pathint_variableschange}
		&&x_e^{(m)} = (P_e x_e-x_e)\frac{m}{M}+x_e + q^{(m)}_e
		-\frac{ (M-m) \xi_e}{2M}+\frac{ m P_e\xi_e }{2M} \,
		\nonumber\\ &&
		x_h^{(m)} = (P_h x_h-x_h)\frac{m}{M}+x_h + q^{(m)}_h
		-\frac{ (M-m) \xi_h}{2M}+\frac{ m P_h\xi_h }{2M} \,
	\end{eqnarray}
	we obtain 
	\begin{eqnarray}\label{pathint_wignerfunctionint2}
		&&W(p,x;\beta) =
		\frac{C(M)}{Z(\beta)N_e!N_h! \lambda_e^{3N_e} \lambda_h^{3N_h}  }  \sum_{\sigma}\sum_{P_e} \sum_{P_h}(\pm 1)^{\kappa_{P_e}+
			\kappa_{P_h}} {\cal S}(\sigma, {\hat P_e}{\hat P_h} \sigma^\prime)
		\big|_{\sigma'=\sigma}\,
		\nonumber    \\&&
		\int d \xi \,
		%\nonumber    \\&&
		\int   dq^{(1)} \dots dq^{(M-1)}
		%    W(p,x;\beta) = Z(\beta)^{-1}
		%    C(M) \frac{1}{\lambda^M}
		%    \int{d\xi}
		%    \int{dx_{1} \dots dx_{M-1}}
		\exp\Biggl\{-\pi \frac{\langle \xi|P_eP_h+E |\xi \rangle}
		{2M } + i\langle \xi |p\rangle
		-\pi \frac{|P_eP_h x-x|^2}{M}
		\nonumber\\&&
		-\sum\limits_{m = 0}^{M-1}
		\biggl[\pi | q^{(m)} - q^{(m+1)}|^2   +
		\epsilon U\biggl((P_eP_h x -x)\frac{m}{M}+x + q^{(m)}
		-\frac{ (M-m) \xi}{2M}+\frac{ m P_eP_h\xi}{2M} \biggr)
		%\Biggr|_{y_0 = y_{M_1} =0 } % x}
		%    \nonumber\\&&
		\biggr]
		\Biggr\}
		%    \Biggr|_{\tilde{y}_{0} = \tilde{y}_{M_2} = 0} %\tilde{x}}.
		\nonumber \\
	\end{eqnarray}
	where angle brackets in $\langle p|\xi\rangle$ mean the scalar product of vectors $ |p \rangle $ 
	and $ |\xi\rangle $, $E$ is unit matrix, while  presenting permutation matrix $P_eP_h$ 
	is equal to unit matrix with appropriately transposed columns. 
	Here and further we imply that momentum and coordinate are dimensionless variables like 
	$p_{l,a} \tilde{\lambda}_a/ \hbar$ and $x_{l,a}/ \tilde{\lambda}_a$, where 
	$\tilde{\lambda}_a=\sqrt{\frac{2\pi\hbar\beta}{m_a M}}$. % is the high temperature thermal wave length. 
	Here constant $C(M)$ as will be shown further is canceled in calculations
	of average values of operators. 
	%Symbolic thermal wave lengths $\tilde{\lambda}=\sqrt{\frac{2\pi\hbar\beta}{mM}}$ without
	%index in denominators means that electron and hole coordinates
	%are devided on the related thermal wave lengths of electrons
	%($\tilde{\lambda_e}$) or holes ($\tilde{\lambda_h}$).
	%
	%For calculation of $\langle \hat A\rangle $ we are going to
	%use the harmonic approximation of the Wigner function in path integral
	%representation (\ref{harmapprox_wigfunctpathintegralfinal}).
	As a result, we have a new representation of Wigner function for
	canonical ensemble in the finite difference form of path integral. 
	Let us note that integration here relates to the integration over
	the Wiener measure of all closed trajectories $\{q^{(0)}, \dots, q^{(M-1)}\}$, which start and end at $q^{(0)}=q^{(M)}=x$.
	In fact, a particle is presented by the closed trajectory with
	characteristic size of order $\lambda_a=\sqrt{\frac{2\pi\hbar\beta}{m_a}}$ in coordinate space. This is
	manifestation of the uncertainty principle.
	
	Then the Wigner function can be written in the following form:
	\begin{eqnarray}
		\label{pathint_wignerfunctionint3}
		&&W(p,x;\beta) =
		\frac{C(M)}{Z(\beta)N_e!N_h!\lambda_e^{3N_e} \lambda_h^{3N_h}} \sum_{\sigma}\sum_{P_e} \sum_{P_h}(\pm 1)^{\kappa_{P_e}+
			\kappa_{P_h}} {\cal S}(\sigma, {\hat P_e}{\hat P_h} \sigma^\prime)
		\big|_{\sigma'=\sigma}\,
		\nonumber \\&&
		\times \exp\Biggl\{-\pi \frac{|P_eP_h x-x|^2}{M} \Biggr\}
		%\nonumber \\&&
		\int dq^{(1)} \dots dq^{(M-1)}
		\exp\Biggl\{
		-\sum\limits_{m = 0}^{M-1}
		\biggl[
		\pi |  q^{(m)}- q^{(m+1)}|^2
		\biggr]
		\Biggr\}
		\nonumber\\&&
		\times \int d \xi \,
		%%\times \int\frac{d \xi }{(2\pi)^{6N}}\,
		\exp\Biggl\{-\pi
		\frac{\langle \xi|P_eP_h+E |\xi \rangle}{2M }
		% \nonumber\\&&
		+ i \Biggl \langle \xi \Bigg| p
		\Biggr \rangle
		\Biggr\}.
	\end{eqnarray}

	In this paper we are going to allow for the exchange effects in
	the pair approximation by effective pseudopotental like have been discussed above (see (\ref{rho-fermi})).
	So in this approximation Wigner function can be written as:
	%this expression is determined by path integral over all closed trajectories,
	%which start and end in the origin $x$, so
	\begin{eqnarray}
		\label{pathint_wignerfunctionint4}
		&&W(p,x;\beta) \approx\,
		\frac{C(M)}{Z(\beta)N_e!N_h!\lambda_e^{3N_e} \lambda_h^{3N_h}}
		%\nonumber \\&&
		%\sum_{P_e} \sum_{P_h}(\pm 1)^{\kappa_{P_e}+
		% \kappa_{P_h}} {\cal S}(\sigma, {\hat P_e}{\hat P_h} \sigma^\prime)
		%\big|_{\sigma'=\sigma}\,
		%\nonumber \\&&
		% \exp\Biggl\{-\pi \frac{|P_eP_h x-x|^2}{M \tilde{\lambda}^2} \Biggr\}
		%\nonumber \\&&
		\int dq^{(1)} \dots dq^{(M-1)}
		\nonumber\\&&
		\times \exp\Biggl\{
		-\sum\limits_{m = 0}^{M-1}
		\biggl[
		\pi | q_{m}-q_{m+1}|^2 \biggr] \Biggr\} \exp\Biggl\{-\frac{M}{4 \pi}
		|p|^2\Biggr\}
		\nonumber\\&&
		\times \sum_{\sigma_e}\Biggl\{1-\sum_{l<t}\delta_{\sigma_{l,e}\sigma_{t,e}}
		\exp(-2\pi \frac{|x_{l,e}-x_{t,e}|^2}{M}
		\delta \biggl(\frac{(p_{l,e}-p_{t,e})\sqrt{M}}{2\pi}
		\biggr)\Biggr\}
		\nonumber\\&&
		\times\sum_{\sigma_h}\Biggl\{1-\sum_{l<t}\delta_{\sigma_{l,h}\sigma_{t,h}}\exp(-2\pi\frac{|x_{l,h}-x_{t,h}|^2}{M} )
		\delta \biggl(\frac{(p_{l,h}-p_{t,h})\sqrt{M}}{2\pi}
		\biggr)\Biggr\}. 
		% \nonumber\\&&
	\end{eqnarray}
	%%where $\lambda_a^2=\frac{2\pi\hbar^2\beta}{m_a}$.
	
	The main idea of deriving expression (\ref{pathint_wignerfunctionint4})
	%and (\ref{pathint_wignerfunctionint5})
	can be explained on example of two
	electrons in 1D space.
	For two electrons the sum over permutations
	consist of two terms
	related to identical permutation (matrix $P$ is equal to unit matrix $E$) and
	non identical permutation (matrix $P$ is equal to matrix $E$ with
	transposed columns).
	To do integration in (\ref{pathint_wignerfunctionint3}) over $\xi$ let us analyze
	eigenvalues of matrix $P+E$. For idetical permutation the eigenvalues are equal to each other and
	are equal to two, while the eigenvalues of matrix $P+E$ related
	to non identical permutation are equal to zero and two.
	Integration over $\xi$ for identical permutation is trivial, while for non identical permutation matrix $P+E$ before integration
	have to be presented in the form $P+E=ODO^{-1}$, where
	$D$ is diagonal matrix with zero and two as the diagonal elements.
	Here matrix $O$ and inverse matrix $O^{-1}$ are given by the formulas:
	\[O = \left|
	\begin{array}{cc}
	1 & 1 \\
	-1 & 1 \\
	\end{array} \right|.\]
	\[O^{-1} = \frac{1}{2}\left| \begin{array}{cc}
	1 &-1 \\
	1 & 1 \\
	\end{array}\right|.\]
	Replacing variables by relation $|\zeta, \eta>=|O^{-1}|\xi>$ for each pair $(l,t)$ we can
	obtain expression (\ref{pathint_wignerfunctionint4}).
	
	To obtain the final expression %(\ref{pathint_wignerfunctionint5})
	we have to approximate delta-function in (\ref{pathint_wignerfunctionint4}) by the standard Gaussian exponent with
	small parameter $\alpha $:
	%and then to change and rescale by thermal wavelength the momentum variables.
	%Replacing delta functions by gaussians in (\ref{pathint_wignerfunctionint4}) we can obtain the final expression in the form:
	\begin{eqnarray}
		\label{pathint_wignerfunctionint5}
		&&W(p,x;\beta) \approx\,
		\frac{C(M)}{Z(\beta)N_e!N_h!\lambda_e^{3N_e} \lambda_h^{3N_h}} \int dq^{(1)} \dots dq^{(M-1)}
		\exp\Biggl\{
		-\sum\limits_{m = 0}^{M-1}
		\biggl[
		\pi |  q^{(m)}- q^{(m+1)}|^2
		\biggr] \Biggr\}
		\nonumber\\&&
		\times \exp\Biggl\{-\frac{M}{4 \pi}
		| p
		% \Biggr|_{x + q^{(m)}}
		|^2\Biggr\}
		\sum_{\sigma}\exp (-\beta \sum_{l<t}v^{e}_{lt})\exp (-\beta \sum_{l<t}v^{h}_{lt})
		% \nonumber\\&&
		% \Biggr\}
		% \nonumber\\&&
		% \Biggr|_{x + q^{(m)}}
		\Biggr|^2\Biggr\}
		% \times \exp\Biggl\{\frac{1}{4 \pi}
		%\Biggr| \frac{i \lambda p}{\hbar} +
		%\frac{\epsilon}{2}\sum\limits_{m = 0}^{M-1} \frac{ (M-2m) }{M}
		% \frac{\partial U(x)}{\partial x}
		% \Biggr|_{x + q^{(m)}} \Biggr|^2\Biggr\}
	\end{eqnarray}
	where
	\begin{eqnarray}
		v^{a}_{lt}\approx-kT \ln\Biggl\{1-\delta_{\sigma_{l,a}\sigma_{t,a}}
		\exp \biggl(-2\pi \frac{|x_{l,a}-x_{t,a}|^2}{M} \biggr)
		\frac{\sqrt{M}}{2\pi  \alpha}\exp \biggl(-\pi\frac{M |p_{l,a}-p_{t,a}|^2}
		{(2\pi \alpha)^2}\biggr)\Biggr\}
		\nonumber
	\end{eqnarray}
	Note that the expression (\ref{pathint_wignerfunctionint5})
	contains explicitly term related the classical Maxwell distribution.
	The others terms account for the influence of
	exchange interaction on the momentum distribution function.
	In the limit of small $\alpha$ the rescaling $p$ by factor $\frac{\sqrt{M}}{2\pi \alpha}$
	regularizes integration over momenta in (\ref{wigfunc_average0})
	and allows to use simplified version of effective pair pseudopotential ($\pi$ is included in small $\alpha^2$):
	\begin{eqnarray}
		v^{a}_{lt}\approx\,-kT \ln\Biggl\{1-\delta_{\sigma_{l,a}\sigma_{t,a}}
		\exp \biggl(-\frac{2\pi|x_{l,a}-x_{t,a}|^2}{\lambda^2_a}\biggr)
		\exp \biggl(-\frac{|(p_{l,a}-p_{t,a})\lambda_a|^2}
		{(2\pi\hbar)^2\alpha^2}\biggr)\Biggr\}
		\nonumber
	\end{eqnarray}
	Momenta and coordinates are written here in natural units ($\lambda_a^2=\frac{2\pi\hbar^2 \beta}{m_a}$).
	%---
	%---
	
	\section{Average values of quantum operators}
	
	For calculation of average values of quantum operators
	$\langle \hat A\rangle $ we are going to
	use the Monte Carlo method (MC) \cite{Metropolis,Hasting}.
	To do this we have to use expression (\ref{pathint_wignerfunctionint5}) presenting the discrete form of path integrals.
	As a result we obtain final expressions for MC calculations in the following form:
	\begin{eqnarray}\label{mmc_averages}
		\langle \hat A\rangle = \frac{\left\langle A(p,x) \right\rangle _w}
		{\langle 1 \rangle _w}.
	\end{eqnarray}
	Here brackets $\bigl\langle g(p,x,q_{1},\dots,q_{M-1})\bigr\rangle _w$ denote
	averaging of any function $g(p,x,q_{1},\dots,q_{M-1})$ with positive
	weight $w(p,x,q_{1},\dots,q_{M-1})$:
	\begin{eqnarray}\label{mmc_averagesav}
		&&\langle g(p,x,q_{1},\dots,q_{M-1}) \rangle _w %=
		\nonumber\\&&
		=\int{dp \rmd x}\int{dq_{1} \dots dq_{M-1}}
		g(p,x,q_{1},\dots,q_{M-1})
		w(p,x,q_{1},\dots,q_{M-1}).
	\end{eqnarray}
	while
	\begin{eqnarray}\label{mmc_averageswfh}
		&&w(p,x,q_{1},\dots,q_{M-1}) =
		\exp\Biggl\{
		-\sum\limits_{m = 0}^{M-1}
		\biggl[
		\pi |  q^{(m)}- q^{(m+1)}|^2
		\biggr] \Biggr\}
		% \nonumber\\
		% \times
		\exp\Biggl\{-\frac{M}{4 \pi}
		\Bigg| p\Bigg|^2
		\biggr) \Biggr\},
		% \nonumber\\&&
		%h(p,x,q_{1},\dots,q_{M-1}) = sign \Biggr(\cos\Biggr\{
		% 2 \frac{1}{4 \pi}
		%\Biggl \langle \frac{ \lambda p}{\hbar} \Bigg|
		%\frac{\epsilon}{2}\sum\limits_{m = 0}^{M-1} \frac{ (M-2m) }{M}
		% \frac{\partial U(x+q^{(m)})}{\partial x}
		%% \Biggr|_{x + q^{(m)}}
		%\Biggr \rangle \Biggr\} \Biggl)
		\nonumber\\&&
		\times \sum_{\sigma}\exp (-\beta \sum_{l<t}v^{e}_{lt})\exp (-\beta \sum_{l<t}v^{h}_{lt})
	\end{eqnarray}
	Note that denominator in (\ref{mmc_averages}) is equal to nominator with
	$A(p,x) = 1$, so
	$C(M)$ in (\ref{pathint_wignerfunctionint5})
	is canceled.

	Calculations of the average values of quantum operators depending only on coordinates of particles 
	is more convenient and reasonable to carry out in configurational space  by standard path integral 
	Monte Karlo method (PIMC).  Within considered above approach it can be done  if we change 
	the following functions: 
	\begin{eqnarray}\label{mc_averageswfh}
		&&\tilde{w}(x,q^{(1)},\dots,q^{(M-1)}) =
		\exp\Biggl\{-\sum\limits_{m = 0}^{M-1} \biggl[
		\pi | q^{(m)} - q^{(m+1)}|^2   +
		\epsilon U(x + q^{(m)}   ) \biggr]   \Biggr\}
		\nonumber\\ &&
		\times \sum_{\sigma}\exp (-\beta \sum_{l<t}\tilde{v}^{e}_{ lt})\exp (-\beta \sum_{l<t}\tilde{v}^{h}_{ lt}) ,
		\nonumber\\ &&
		\tilde{h}(x,q^{(1)},\dots,q^{(M-1)}) \equiv 1 .
	\end{eqnarray}
	where $\tilde{v}^{a}_{ lt}$ is defined by equation~(\ref{pairfermi}),  
	\begin{eqnarray}\label{mcg_averagesav}
		\langle  g(x,q^{(1)},\dots,q^{(M-1)}) \rangle_{\tilde{w}}=
		\int{dx}\int{dq^{(1)} \dots dq^{(M-1)}}
		g(x,q^{(1)},\dots,q^{(M-1)})
		\tilde{w}(x,q^{(1)},\dots,q^{(M-1)})
	\end{eqnarray}
	and 
	\begin{eqnarray}\label{mm_averages}
		\langle  {\tilde{ \hat A}}\rangle  =
		\frac{\left\langle \tilde{A}(x)
			\cdot \tilde{h}(x,q^{(1)},\dots,q^{(M-1)})\right\rangle_{\tilde{w}}}
		{\langle
			\tilde{h}(x,q^{(1)},\dots,q^{(M-1)})\rangle_{\tilde{w}}}.
	\end{eqnarray}
	%only for overall positions $q$ while trajectories $q(\tau)$ representing particles
	%%
%%	\newpage
	\section{Results of numerical calculations}
	We define momentum distribution functions and pair correlation functions  
	for holes ($a=h$) and electrons ($a=e$) by the following expressions:
	\begin{eqnarray}  \label{gab-rho}
		&& w_a(|p|) =  
		\frac{\left\langle \delta(|p_{1,a}|-|p|) 
			\cdot h(p,x,q^{(1)},\dots,q^{(M-1)})\right\rangle _w}
		{\langle
			h(p,x,q^{(1)},\dots,q^{(M-1)})\rangle_w} 
		\nonumber\\&&   
		g_{ab}(r)  =
		\frac{\left\langle \delta(|x_{1,a}-x_{1,b}|-r) 
			\cdot \tilde{h}(x,q^{(1)},\dots,q^{(M-1)})\right\rangle_{\tilde{w}}}
		{\langle
			\tilde{h}(x,q^{(1)},\dots,q^{(M-1)})\rangle_{\tilde{w}}} 
		%	&&w_a(|p|) = \int\limits_{V}\frac{dpdx}{(2\pi)^{6N}} \,\delta(|p_{1,a}|-|p|)\,
		%W(p,x;\beta) \,
		%\nonumber\\ &&
		%g_{ab}(r) = \int\limits_{V} dx \,\delta(|x_{1,a}-x_{1,b}|-r)\,
		%\tilde{W}(x;\beta) \,
	\end{eqnarray}
	where $\delta$ is delta function, $a$ and $b$ are types of the particles.  
	
	To test the developed approach we have carried out calculations of the path integral  
	representation of Wigner function in the form (\ref{pathint_wignerfunctionint5}). 
	To extent the region of applicability of pair approximation we have 
	used the small parameter $\alpha^2$  as adjustable function  of the universal degeneracy parameter of ideal 
	fermions $n\lambda^3$, namely  $\alpha^2_a=0.00505+0.056n\lambda_a^3$ .  
	%with unit matrix $\chi\equiv\widehat{1}$. 
	Calculations have been done for two hundred particles each presented by 
	twenty beads. Results have been obtained by averaging-out over one million 
	particle configurations. To simplify calculations we fix the number of electrons and holes with 
	the same spin projection equal to $N_e/2$ and $N_h/2$ respectively. 
	
		%\newpage
		\begin{figure}[htp]
			%\vspace{0cm} \hspace{0.0cm}
			\includegraphics[width=7cm,clip=true]{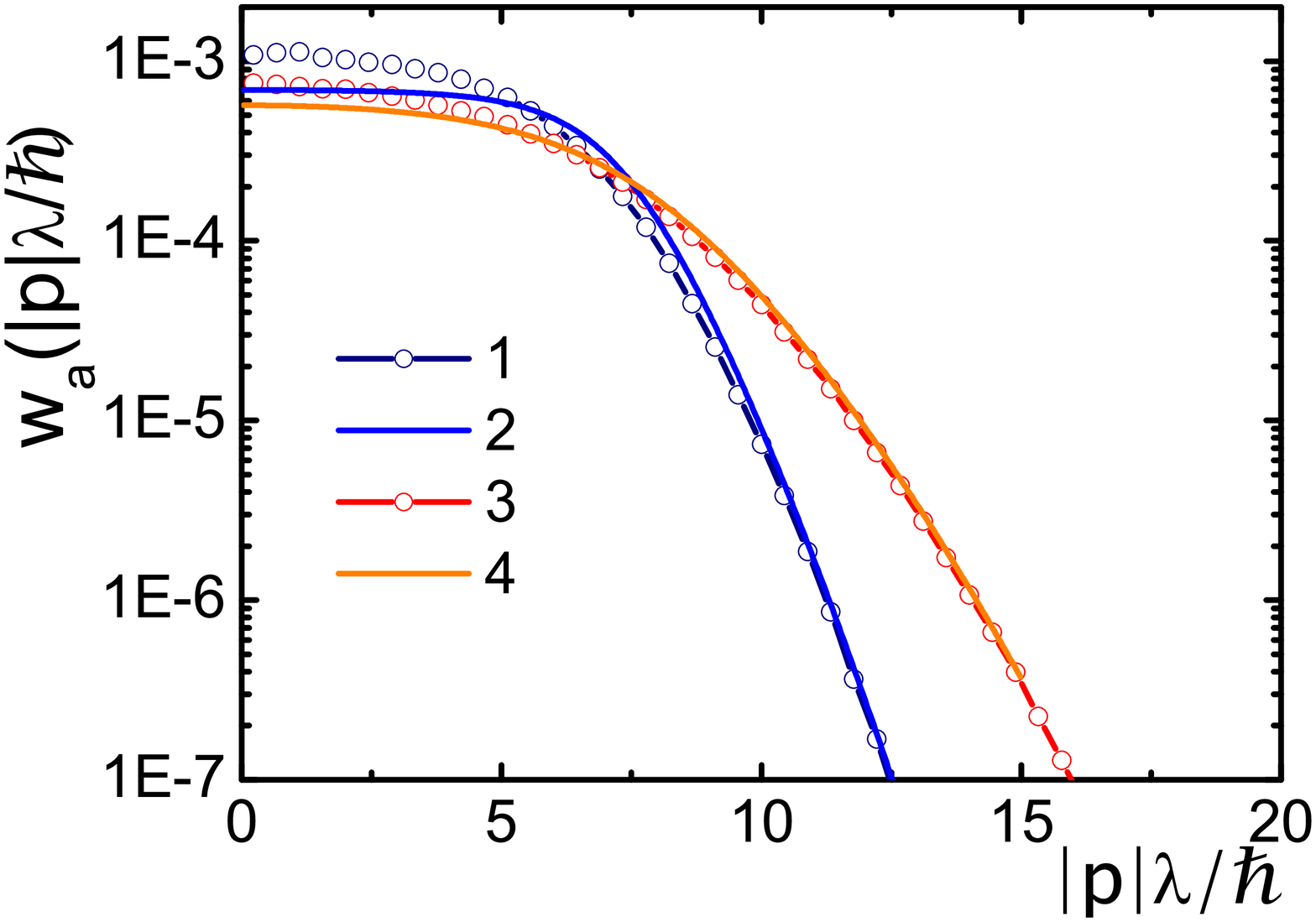}
			\includegraphics[width=7cm,clip=true]{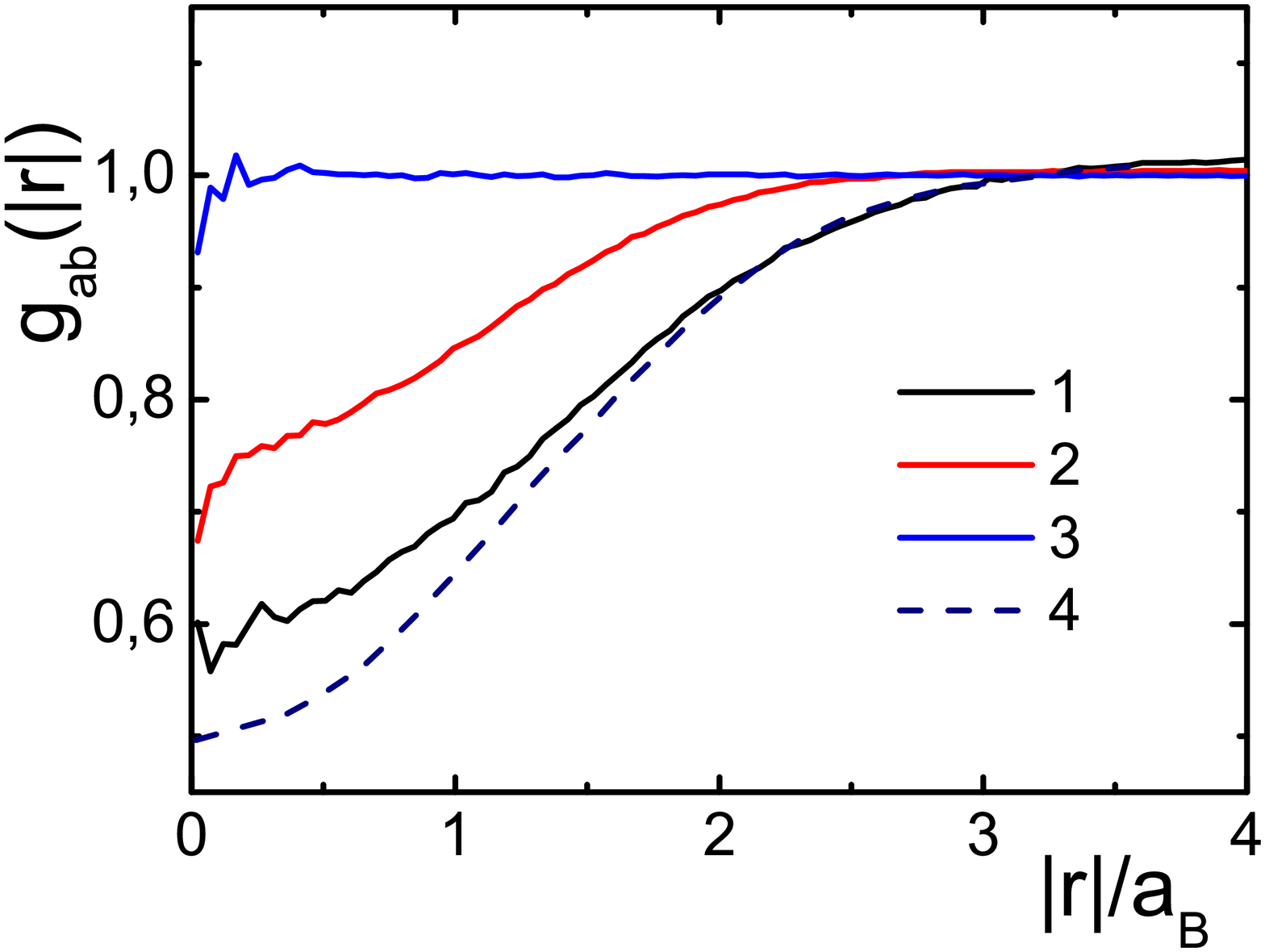}
			\includegraphics[width=7cm,clip=true]{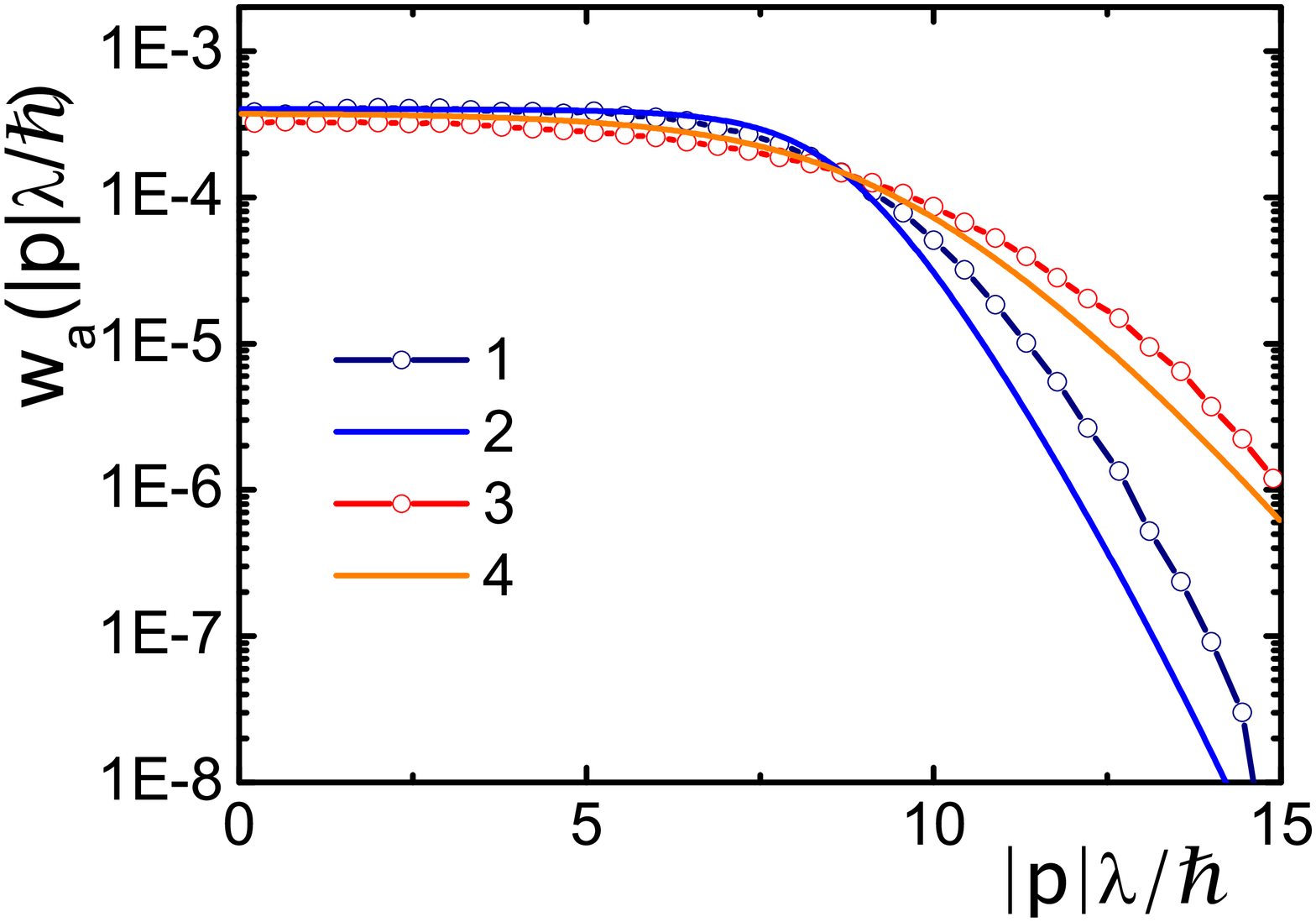}
			\hspace{1.7cm}			
			\includegraphics[width=7.00cm,clip=true]{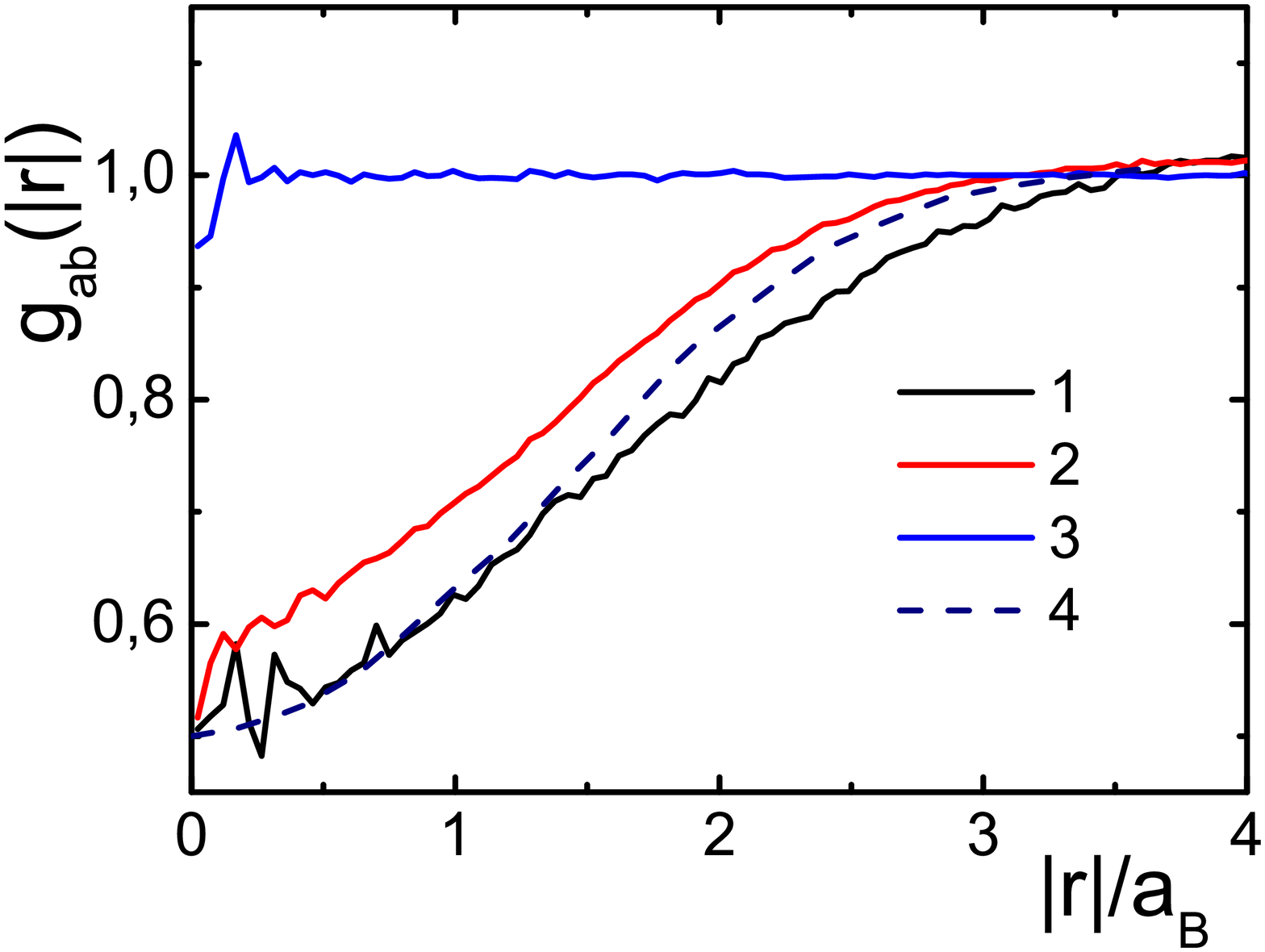}
			\includegraphics[width=7.00cm,clip=true]{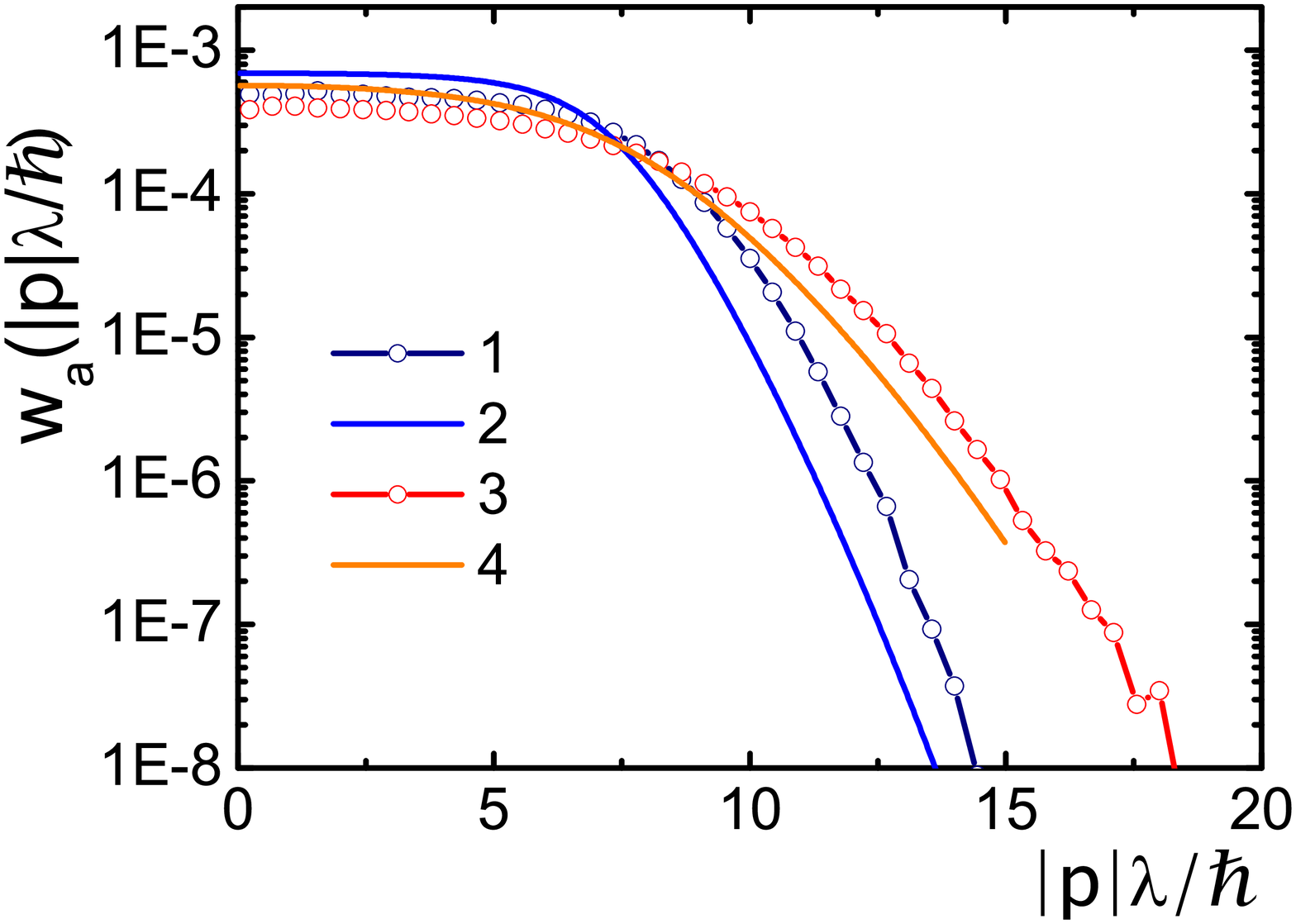}
			\includegraphics[width=7.00cm,clip=true]{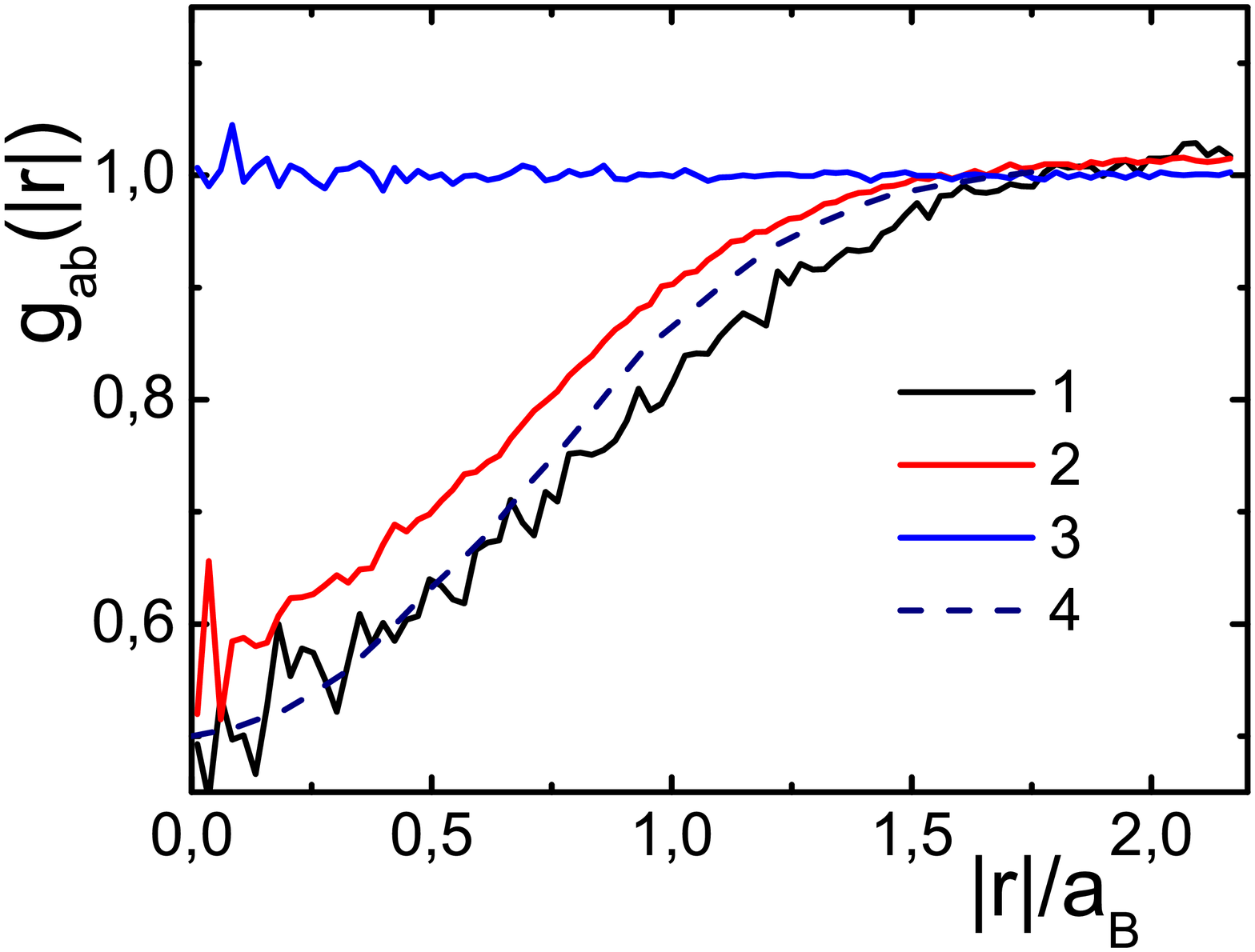}
			\includegraphics[width=7.00cm,clip=true]{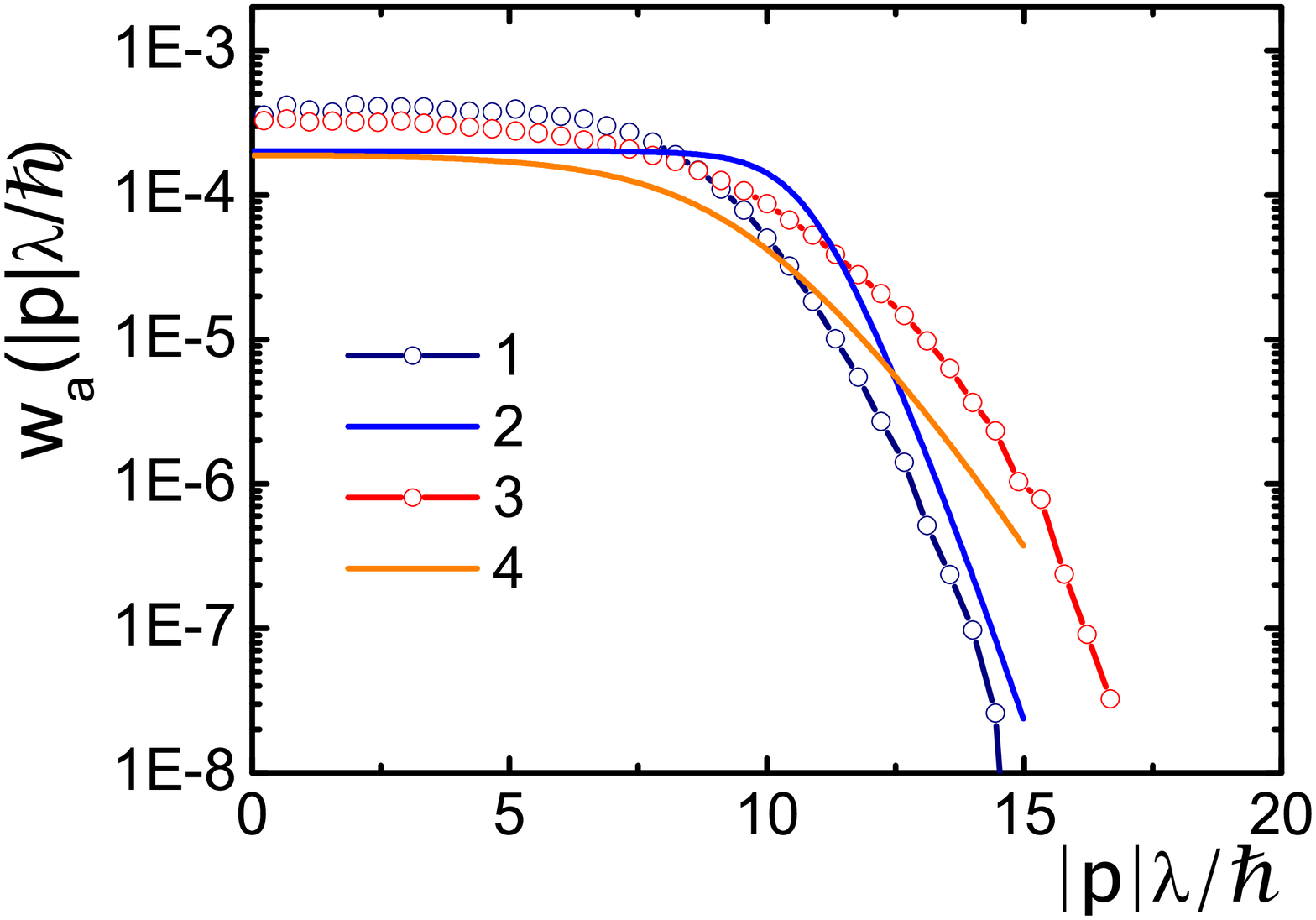}
			\hspace{1.7cm}
			\includegraphics[width=7.00cm,clip=true]{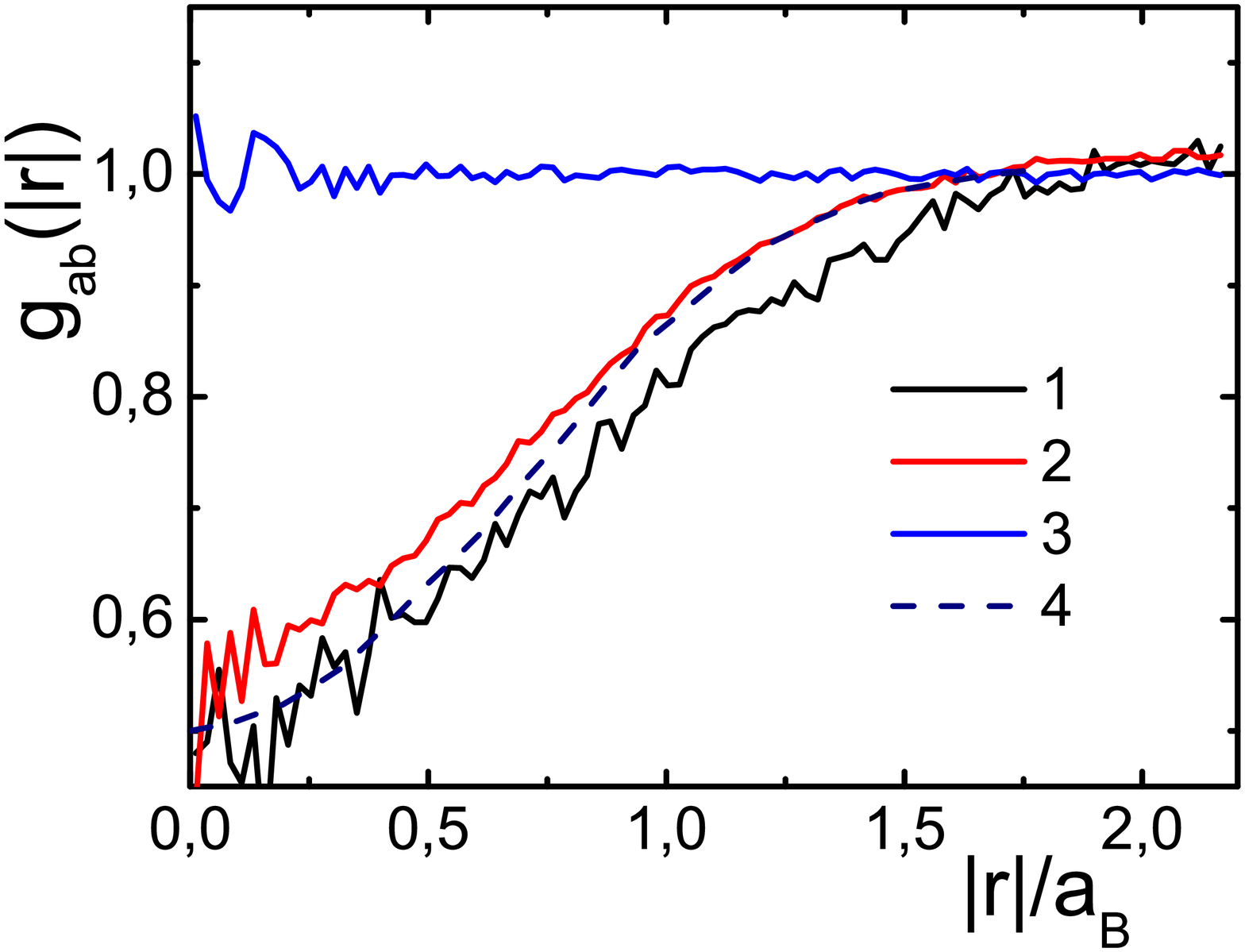}
			%\hspace{1.8cm}
			%\hspace{1.8cm}
			\caption{(Color online) The momentum distribution functions $w_a(|p|)$ (left panels) and pair correlation functions 
				$g_{ab}(|r|),(a,b=e,h)$ (right panels) for ideal electron - hole plasma. % at $r_s=2$. 
				Left panels:
				lines $1,3$ show PIMC distributions $w_a(|p|)$  scaled by ratio of Plank constant to the electron thermal wavelength
				($\frac{\hbar}{\lambda_e}$), while lines $2,4$ demonstrate the ideal Fermi distributions for electrons and two times heavier  holes respectively;
				Right panels:
				lines $1,2,3$ present PIMC electron -- electron, hole -- hole and electron -- hole  correlation functions scaled by Bohr radius respectively, 
				while line 4 show results of analytical approximations \cite{Bosse} for electrons. Parameters of degeneracy $n \lambda^3_e$ for
				electrons are increasing from upper to bottom rows as $n \lambda^3_e = 5,10,15,20$.
			}
			\label{DistrFunc}
		\end{figure}

	For ideal electron--hole plasma Figure~\ref{DistrFunc} shows
	the momentum distributions $w_{a}(|p_a|),(a=e,h)$ and pair correlation functions $g_{ab}(|r|),(a,b=e,h)$ scaled by ratio of the Plank constant to the electron thermal wavelength ($\frac{\hbar}{\lambda_e}$) and Bohr radius ($a_B$) respectively. In left column of Figure~\ref{DistrFunc} results of Monte Carlo calculations for electrons and holes are presented by lines 1 and 3, while lines 2 and 4 shows ideal Fermi distributions. Presented distribution functions are normalized to one.
%	Parameters of degeneracy $n\lambda_e^3$ for electrons are increasing from upper to bottom rows as $n\lambda_e^3=5,10, 15, 20$. 
	Let us note that holes is these calculations are two times heavier than electrons, so the related parameters of degeneracy is $2^{3/2}$ times smaller.
	As it follows from the analysis of Figure~\ref{DistrFunc} agreement of PIMC calculations and analytical Fermi distribution are good enough up to parameter of degeneracy equal to $n\lambda_e^3=15$.
	%%Comparison of the ideal Fermi distributions with results of our calculations
	%%for ideal plasma demonstrate good accuracy of our approach in wide range of
	%%particle momentum, in which decay of the distribution functions is about of five orders of magnitude.
	
	It necessary to stress that one of the reason of increasing discrepancy at large degeneracy of femions is limitation on
	available computing power allowing calculations
	with several hundred particles in Monte Carlo cell. When parameter of degeneracy is approaching $20$
	the thermal electron wave length is of order Monte Carlo cell size and influence finite number of particles and
	periodic boundary conditions becomes significant as was tested by our calculations.
	
	Right column of Figure~\ref{DistrFunc} presents results of Monte Carlo calculations of pair correlation functions $g_{ab}(|r|),(a,b=e,h)$.
	%Lines 1.2.3 present electron---electron, hole---hole and electron---hole distribution functions respectively.
	Influence of Fermi repulsion at distance less than thermal 
	wave length is evident enough. 
	%  with increasing parameter of degeneracy from upper to bottom rows. 
	At the same time the electron---hole pair correlation 
	functions are identically equal to one as exchange interaction between particle of different type is missing.

	% \newpage
	\begin{figure}[htb]
		%\vspace{0cm} \hspace{0.0cm}
		\includegraphics[width=7.25cm,clip=true]{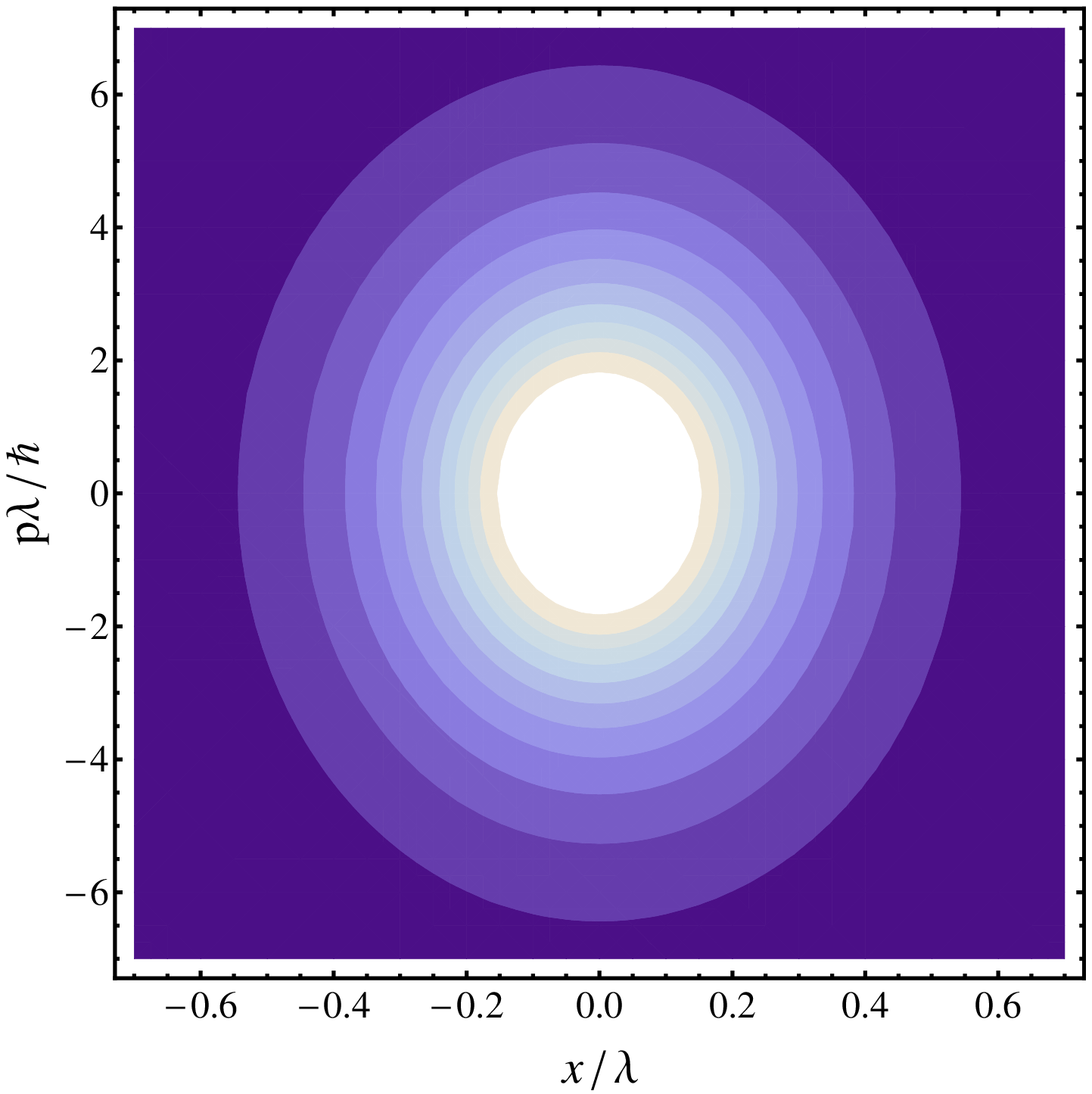}
		\hspace{1.3cm}
		\includegraphics[width=7.25cm,clip=true]{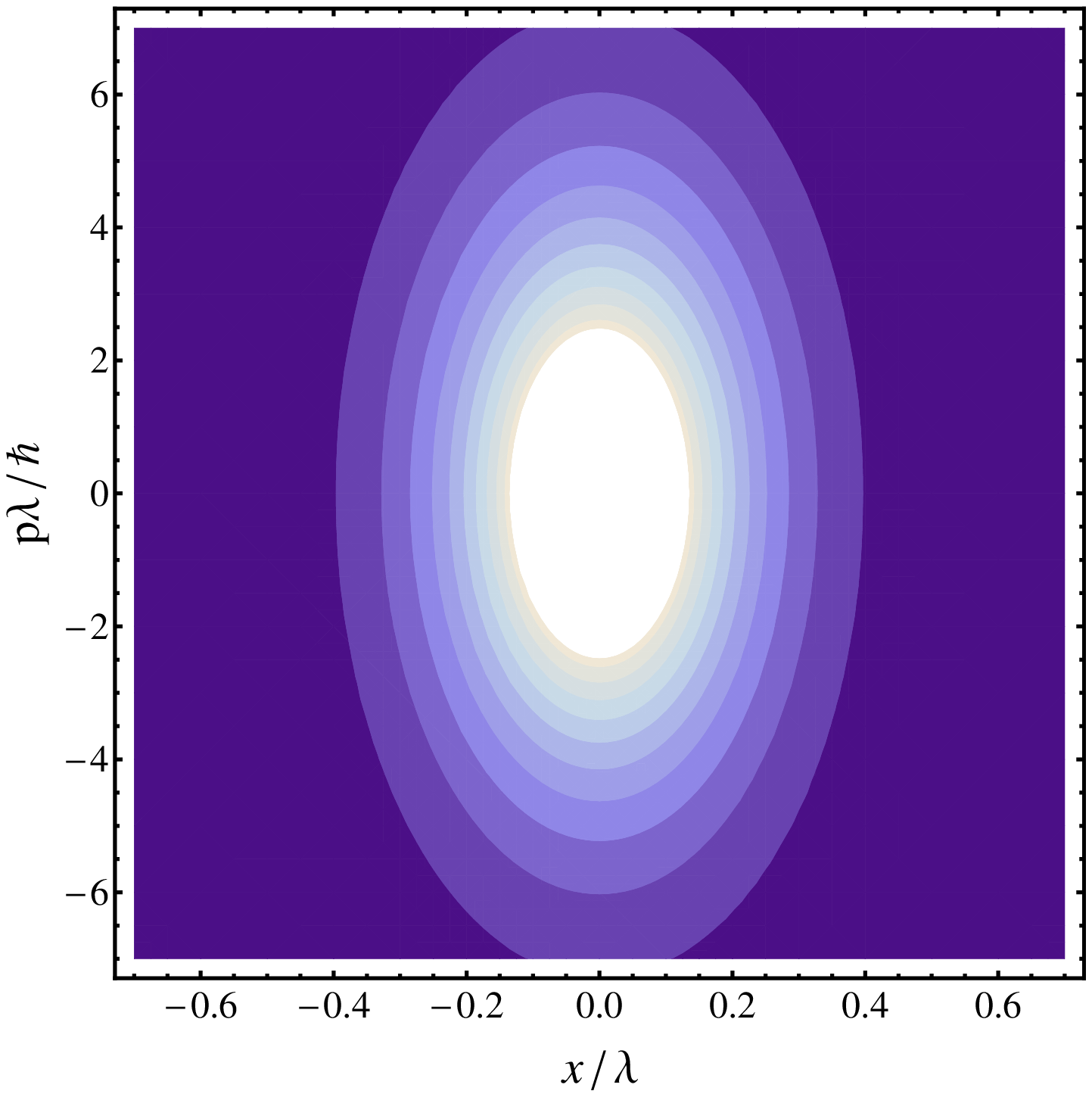}
		%\hspace{1.8cm}
		%\hspace{1.8cm}
		\caption{(Color online) Contour plots of the repulsive effective exchange pair pseudopotentials in phase space.
			Left panel:  dark area $\beta v^{e}_{lt}\approx 0$, white area $\beta v^{e}_{lt}\geq 1.9$ ,  $m_e=1$.
			Right panel: dark area $\beta v^{h}_{lt}\approx 0$, white area $\beta v^{h}_{lt} \geq 1.5$ , $m_h=2$. 
		}
		\label{ExPt}
	\end{figure}
	
	Presented results have been obtained in pair exchange approximation described by introduced above the effective pair pseudopotentials.
	Figure~\ref{ExPt} presents contour plots of exchange pair pseudopotentials for parameter of degeneracy equal to $5.6$.
	Momenta and coordinates axises
	are scaled by the electron thermal wave length with Plank constant and factor ten for momentum. As before holes are two times heavier than electrons.
	As it follows from analysis of Figure~\ref{DistrFunc} the Pauli blocking of fermions in phase space accounting for by these exchange pseudopotentials
	provides agreement of PIMC calculations and analytical Fermi distribution in wide ranges of fermion degeneracy and fermion momenta,
	where decay of the distribution functions is about of five orders of magnitude.
	
	\section{Conclusion}
	
	The new path integral representation of the quantum Wigner function in
	the phase space has been developed for canonical ensemble. Explicit analytical expression of the Wigner function accounting for Fermi statistical effects by effective pair pseudopotential has been obtained. Derived pseudopotential depends
	on coordinates, momenta and degeneracy parameter of fermions. The new quantum Monte-Carlo method for calculations 
	of average values of arbitrary quantum operators has been proposed. 	To test the developed approach calculations 
	of the momentum distribution function and pair correlation functions of the ideal system of Fermi particles has been carried out.
	Calculated by Monte Carlo method the momentum distributions for degenerate ideal fermions are in a good agreement 
	with analytical Fermi distribution
	in a wide range of momentum and degeneracy parameter.
	Generalization of this approach for studies influence of interparticle interaction on momentum distribution functions of strongly coupled 
	Fermi system is in progress. First results show appearance of long quantum 'tails' in the Fermi distribution functions. 
	
	%---\section*{A
	%%\section{Acknowledgements}
	\ack
	We acknowledge stimulating discussions with Profs. A.N.~Starostin, Yu.V.~Petrushevich, E.E.~Son, I.L.~Iosilevskii, M.~Bonitz and V.I.~Man'ko.
	%G.~Kalman, P.~Levai, D.~Blaschke, R. Bock and H.~Stoecker.
	%Y.B.I. was partially supported by grant of the Russian Ministry of
	%Science and Education NS-215.2012.2.
	%%%%This work has been supported by the Russian Science Foundation via grant 14-50-00124.

\end{document}